\documentclass[conference]{IEEEtran}
\IEEEoverridecommandlockouts
\usepackage{cite}
\usepackage{indentfirst}
\usepackage{multirow}
\usepackage{booktabs}
\usepackage{makecell}
\usepackage{float}
\usepackage{amsmath,amssymb,amsfonts}
\usepackage{algorithmic}
\usepackage{graphicx}
\usepackage{graphicx,subfigure}
\graphicspath{{./}}
\usepackage{textcomp}
\usepackage{xcolor}
\usepackage{subfigure}
\usepackage{booktabs}
\usepackage{bm}
\usepackage{amsthm}
\usepackage{epstopdf}
\def\BibTeX{{\rm B\kern-.05em{\sc i\kern-.025em b}\kern-.08em
		T\kern-.1667em\lower.7ex\hbox{E}\kern-.125emX}}

\begin{document}
	\bibstyle{IEEEtran}
	\title{Agent-Native Task-Oriented Communication with Joint Token Compression Coding and Modulation}
	
	\author{
		\IEEEauthorblockN{Zhuoran~Xiao\IEEEauthorrefmark{1}, Yihang~Huang\IEEEauthorrefmark{2},Tianyu~Jiao\IEEEauthorrefmark{4}, Xiaohua~Xu\IEEEauthorrefmark{3}, and Yin Xu\IEEEauthorrefmark{4}
		}
		
        \IEEEauthorblockA{\IEEEauthorrefmark{1}UNISOC (Shanghai) Technologies Co., Ltd. Shanghai, China \\ \IEEEauthorrefmark{2}Chongqing University of Posts and Telecommunications, Chongqing, China\\ \IEEEauthorrefmark{3}Zhaotong University, Zhaotong, China\\ \IEEEauthorrefmark{4}Shanghai Jiao Tong University, Shanghai, China\\
	E-mail: Zhuoran.Xiao@unisoc.com}
		\thanks{Yihang Huang and Xiaohua Xu are co-corresponding authors.}}
		\vspace{-50cm}
	\maketitle

\begin{abstract}
As large foundation models empower agents to become pervasive across industries and emerge as central actors in intelligent systems, a fundamental rethinking of communication paradigms toward AI-native, agent-centric designs in the post-Shannon era becomes inevitable. One essential shift is that tokens, which are the minimal semantic units natively processed by large language models (LLMs), should replace bits as the fundamental unit of communication. However, existing works in the LLMs field assume lossless token transmission over high-speed wired links and largely neglect the air-interface overhead and channel distortions inherent in wireless environments, lacking a native design for wireless token communication systems. To bridge this gap, we propose an innovative design for a token transmitter-receiver architecture that facilitates task-oriented token transmission. Specifically, we propose JTCM (Joint Token Coding and Modulation), an AI-native semantic communication framework that jointly optimizes token representation, channel coding, and modulation to maximize downstream task performance directly. Correspondingly, we propose a two-stage training scheme. In the first stage, the token encoder–decoder pair is pre-trained to enable semantic-preserving compression and reconstruction. In the second stage, it is fine-tuned end-to-end with a multi-modal foundation model under specific downstream tasks to achieve task-aware optimization. Extensive experiments demonstrate that JTCM significantly reduces transmission overhead while enhancing task accuracy and robustness compared to state-of-the-art baselines in bandwidth- and SNR-constrained wireless channels.
\end{abstract}
	
	
\section{Introduction} \label{intro}
The continuous advancement of large language models is catalyzing a fundamental shift across industries. LLM agents, which are powered by multi-modal perception, reasoning, memory, and decision-making capabilities, are increasingly becoming the primary actors in intelligent systems. Consequently, traditional communication paradigms, which are centred on human-to-human or human-to-machine interaction, are shifting toward agent-to-agent communication. This transition calls for a new AI-native communication architecture, which diverges from conventional designs optimized for bit-level fidelity and instead aligns intrinsically with the semantic and task-oriented nature of intelligent agents.
		
The most essential principle underlying agent-centric communication is that tokens, which are the fundamental unit natively processed by LLMs, should serve as the basic unit of transmission. Tokens exhibit several intrinsic properties that fundamentally distinguish them from classical bits, necessitating a rethinking of communication system design. First, tokens convey high-level semantic meaning rather than raw binary information. Thus, the primary objective of token transmission is not exact reconstruction, but semantic interpretability. Second, the meaning of a token is not fixed but inherently contextual, shaped by factors such as the agent’s role, historical interactions, and the dynamic environment. Third, tokens possess native robustness to noise and loss. Semantic integrity is often preserved even when certain tokens are dropped from a sequence or when channel distortions perturb their embedding vectors. These characteristics collectively imply that conventional systems transmitting raw data via bitstreams are fundamentally misaligned and highly inefficient for agent-centric communication scenarios.

Several studies have begun exploring new paradigms of communication for meeting the requirements of semantic convey and task-oriented transmission. DeepJSCC \cite{10328187} introduced a deep learning-based joint source-channel coding framework, which jointly optimizes source representation and channel transmission using neural networks. This paradigm has since been extended to multi-media scenarios, including image and video delivery over wireless channels \cite{9953110}. Recognizing that full data reconstruction is often unnecessary for specific objectives, task-oriented semantic communication frameworks have emerged \cite{10333632}, aiming to transmit only the information essential for downstream task execution. Furthermore, \cite{10570717} harnesses LLMs to perform semantic-aware compression of textual inputs, while \cite{10615340} integrates LLMs with diffusion models to enable high-level semantic transmission of visual content, bypassing pixel-level fidelity.

\begin{figure*}[htb!]
	\centering
	\includegraphics[width=0.65\textwidth]{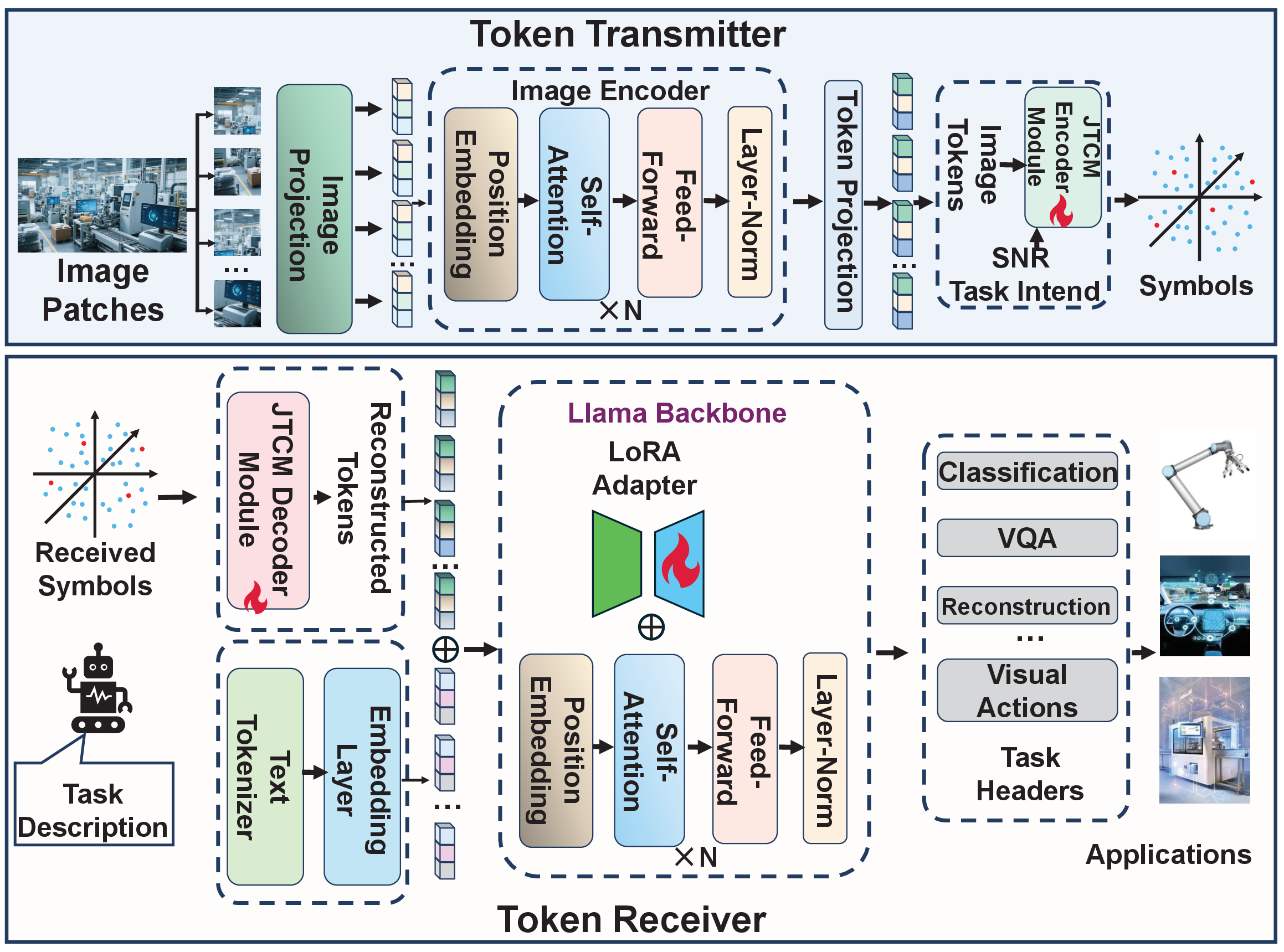}
	\vspace{-.3cm}
	\caption{The system model of the proposed Token transmitter and receiver empowered agent communication.}
	\label{sys_model}
	\vspace{-.5cm}
\end{figure*}	

Despite the above-mentioned works having taken a step ahead of traditional communication architecture, they remain inefficient and fundamentally misaligned with the agent-centric vision. Conventional JSCC schemes prioritize perceptual or pixel-level reconstruction fidelity, optimizing metrics such as PSNR that exhibit weak correlation with downstream task performance. Meanwhile, emerging semantic communication frameworks often treat semantic extraction and physical-layer transmission as decoupled stages. More critically, these approaches typically target a single task type or modality, lacking the generalization capability required for versatile agents. These limitations stem from their underlying design philosophy, which operates at the level of modality-specific signal compression, rather than leveraging tokens as the fundamental units of communication. Consequently, none of these methods directly optimize an end-to-end pipeline from token modulation to task execution, nor do they exploit the intrinsic robustness, context dependence, and sparsity of multimodal tokens under realistic wireless channel constraints. As a result, they incur unnecessary transmission overhead and fail to achieve true integration of communication and computation.
	
To bridge this gap, this paper proposes JTCM (Joint Token Coding and Modulation), an AI-native, task-oriented semantic communication framework tailored for agent-to-agent communication in wireless environments. Generally, we propose a token transmitter network that learns an end-to-end mapping from multi-modal token sequences to non-uniformly distributed constellation symbols for wireless transmission, and a token receiver that directly maps the distorted channel outputs to the final predictions of downstream tasks. We further introduce a two-stage training strategy that firstly pre-trains the transceiver for semantic-preserving token compression, then fine-tunes it jointly with a foundation model under specific downstream objectives. This design is grounded on three key principles. First, exploits the inherent redundancy and robustness of token-level semantics by co-designing compression, coding, and modulation within a unified pipeline.
Second, the entire system is optimized directly for downstream task performance, rather than intermediate signal fidelity. Third, since multi-modal inputs are naturally aligned in the token space of foundation models, JTCM naturally supports diverse modalities and tasks without architectural reconfiguration.

The remainder of this paper is structured as follows. Section \ref{Problem} presents the system model and the task objective. The rationale behind the design of our network and training scheme is discussed in Section \ref{model}. Section \ref{experiments} details the experimental setup and reports the results, providing a comparative evaluation of the proposed approach against existing state-of-the-art (SOTA) methods from multiple perspectives. Finally, Section \ref{conclusion} summarizes the main conclusions of this work.	

\section{System Model And Task Description} \label{Problem}
As illustrated in Fig. \ref{sys_model}, this paper investigates an end-to-end communication system designed for semantic token transmission between two agents. Specifically, Agent-2, which is equipped with a token receiver, requires task-relevant information from Agent-1 to perform a downstream task successfully. Agent-1, which has access to the raw input, encodes this information into a compact sequence of semantic tokens and transmits them over a noisy wireless channel using a learned joint token coding and modulation scheme. 

The transmitted symbols are subject to channel distortions. Upon reception, Agent-2 decodes the corrupted token sequence and uses it to generate the final task output. Crucially, both the encoder at Agent-1 and the decoder at Agent-2 are trained end-to-end to maximize downstream task performance, thereby enabling the system to prioritize semantically meaningful information over perceptual details. This design philosophy aligns with the principles of semantic communication, where the goal is reliable task execution rather than perfect data recovery.

\begin{figure*}[htb!]
	\centering
	\includegraphics[width=0.7\textwidth]{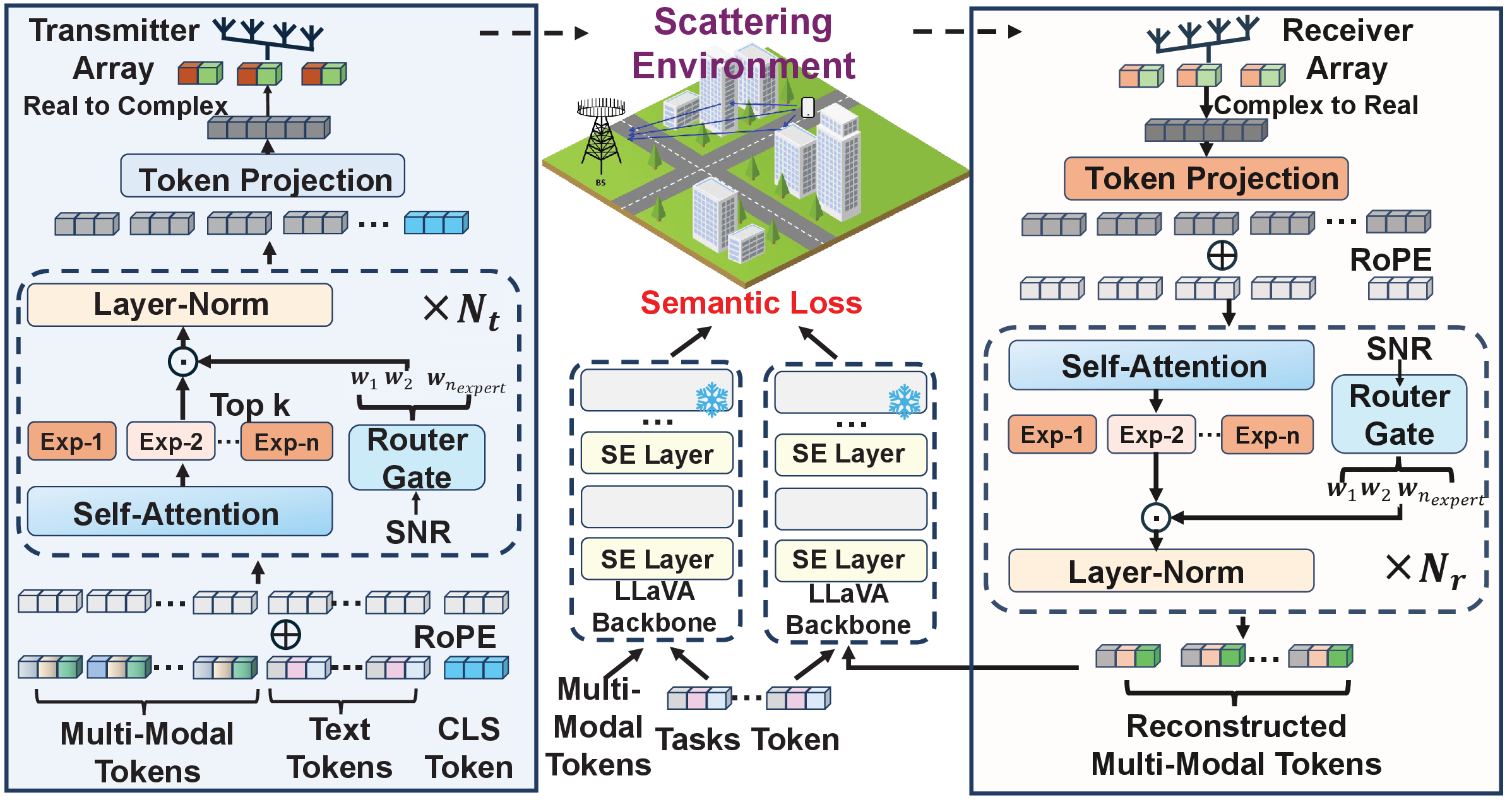}
	\vspace{-.3cm}
	\caption{The training process of the proposed JTCM modules.}
	\label{training_process}
	\vspace{-.5cm}
\end{figure*}

\section{Model Design And Learning Scheme} \label{model}
\subsection{Model Design}
This section describes the architecture of JTCM in the forward direction of information flow shown in Fig. \ref{sys_model} and Fig. \ref{training_process}. It is worth mentioning that we utilize images as the representation of multimodal input. In fact, all kinds of modalities follow the same process.
\subsubsection{Token Transmitter}
The token transmitter is tasked with converting raw multi-modal inputs into a compact, channel-adapted symbol stream that preserves task-relevant semantics. To minimize computational overhead on resource-constrained agents, the transmitter is deliberately kept lightweight and leverages pre-trained components wherever possible.

Given an input image $\mathbf{X}_{img} \in \mathbb{R}^{H \times W \times 3}$, we employ a frozen, pretrained image encoder to map it into a sequence of vision tokens, which can be written as
\begin{equation}
	\mathbf{V} = \mathbf{enc}_{image}(\mathbf{X}_{img}) \in \mathbb{R}^{N \times d}
\end{equation}
where $N$ is the number of tokens and $d$ is the embedding dimension. This encoder has been trained via contrastive learning to align visual representations with textual embeddings in a shared semantic space. Consequently, the resulting visual tokens are natively compatible with multi-modal large language models by being directly concatenated with text tokens and fed into the LLM backbone. 

Then, we prepend the task intention prompt $\mathbf{P}_{intend}$ to the image token sequence, forming the joint input
\begin{equation}
	\mathbf{Z}_{in} = \left [ \mathbf{P}_{intention}, \mathbf{V} \right ] \in \mathbb{R}^{(M+N) \times d}
\end{equation}
where $M$ denotes the number of intention prompt tokens. The compression induced by JTCM can be viewed as a task-directed collapse. This prompt acts as a dynamic prefix that conditions the subsequent compression and modulation process on the downstream objective. 

The JTCM encoder $f_{\theta}$ then maps $\mathbf{Z}_{in}$ to a complex-valued symbol sequence 
\begin{equation}
	\mathbf{s} = f_{\theta}(\mathbf{Z}_{in}) \in \mathbb{C}^L.
\end{equation}
In practice, $f_{\theta}$ outputs a real-valued vector of length $2L$, which is reshaped into $L$ complex symbol sequence is then normalized to satisfy average power constraint $\mathbb{E} \left [ \left | \mathbf{s}_l \right |^2  \right ] = 1$, and mapped onto resource elements of a MIMO-OFDM waveform for over-the-air transmission.

\subsubsection{Token Receiver}
At the receiver side, the distorted symbols $\hat{\mathbf{s}} \in \mathbb{C}^L$ are obtained from the wireless channel. Then, they are first fed into the JTCM decoder module $g_{\phi}$, which generates a sequence of semantic tokens with the same length as the original visual token sequence
\begin{equation}
	\mathbf{\hat{V}} = g_{\phi}(\hat{\mathbf{s}}) \in \mathbb{R}^{N \times d}.
\end{equation}

To enable task-aware reasoning, a detailed natural-language description of the downstream task is tokenized using the frozen tokenizer of the pre-trained LLM. The resulting text tokens are then embedded via the LLM's fixed input embedding layer to obtain the task prompt representation $\mathbf{P}_{task} \in \mathbb{R}^{M' \times d}$, where $M'$ denotes the number of text tokens. The recovered visual tokens $\mathbf{\hat{V}}$ and the embedded task prompt $\mathbf{P}_{task}$ are concatenated to form the joint input sequence
\begin{equation}
	\mathbf{Z}_{rec} = \left [ \mathbf{P}_{task}, \mathbf{\hat{V}} \right ] \in \mathbb{R}^{(M'+N) \times d}.
\end{equation} 
The joint input sequence is subsequently passed through a LoRA-finetuned LLM backbone.

Finally, the system supports two flexible output strategies depending on the application scenario. One is that the original language modeling head (LM head) of the LLM can be retained to generate free-form answers directly. The other is that the language modeling head is replaced by task-specific heads (e.g., a classifier for image recognition), enabling optimized performance on structured tasks. This design ensures that JTCM remains compatible with both open-ended and closed-set agent tasks, while maintaining end-to-end differentiability from channel output to final prediction during training.
\subsubsection{JTCM Modules}
As illustrated in Fig. \ref{training_process}, both the JTCM encoder and decoder are built upon symmetric Mixture-of-Experts (MoE) Transformer architectures. Specifically, within each Transformer block, the standard feed-forward network (FFN) is replaced by an MoE layer consisting of multiple expert sub-networks. Only the top-$k$ experts are activated per token, with routing weights determined by a lightweight router gate.

At the transmitter, the input sequence $\mathbf{Z}_{in}$ is first augmented with a learnable CLS token $\mathbf{z}_{cls} \in \mathbb{R}^d$ at the end to globally aggregate semantic information across all tokens. To overcome the limitation of fixed positional encodings in dynamic wireless scenarios, we employ RoPE (Rotary Position Embedding) throughout the encoder. 

The router gate network takes the instantaneous channel SNR $\gamma$ as input and outputs gating weights $\mathbf{w} = \mathrm{Gate}(\gamma) \in \mathbb{R}^{n_{expert}}$, where $n_{expert}$ is the number of experts. The network design enables the model to dynamically adapt its internal computation based on the scattering environments, enhancing robustness across a wide SNR range without requiring retraining. 

The output CLS token embedding $\mathbf{h}_{cls}$ from the final MoE layer is passed through a token projection head, which is a linear projection layer, to reproduce a real-valued vector of dimension $2L$
\begin{equation}
	\mathbf{u} = \mathrm{Proj}_{tx}(\mathbf{h}_{cls}) \in \mathbb{R}^{2L}.
\end{equation}
This vector is then reshaped into $L$ complex symbols by pairing consecutive elements' real and imaginary parts
\begin{equation}
	s_l = u_{2l-1} + j \cdot u_{2l}, \quad l = 1,\dots,L
\end{equation}

At the receiver, the distorted complex symbols $\mathbf{\hat{s}} \in \mathbb{C}^L$ are first converted to a real vector $\mathbf{v'} \in \mathbb{R}^{2L}$ by concatenating real and imaginary parts $\mathbf{v} =     \left [ \mathbf{Re}(\mathbf{\hat{s}}),\mathbf{Im}(\mathbf{\hat{s}}) \right ] $. This vector is projected back to the token space via a symmetric token projection head $\mathrm{Proj}_{rx}(\cdot)$, yielding an initial token sequence $\hat{\mathbf{Z}}^{(0)} \in \mathbb{R}^{N \times d}$ that matches the length of the original multimodal input. RoPE is again applied, and the sequence is processed through the MoE-based decoder. Notably, the router gate shares the same architecture and parameters as the transmitter’s gate, using the estimated SNR $\gamma$ to ensure consistent expert selection across the link.

\subsection{Training Scheme}
To ensure stable convergence and robust generalization across diverse channel conditions, we adopt a two-stage training strategy. We first pre-train the JTCM modules for semantic-preserving token compression under varying SNRs, followed by task-aware fine-tuning with highly efficient fine-tuning on the foundation model. This staged approach decouples representation learning from task optimization, significantly improving convergence of the training process.
\subsubsection{Pre-Training of JTCM Modules}
During the pre-training stage, both the original multi-modal input token sequence $\mathbf{V}$ and the reconstructed token sequence $\mathbf{\hat{V}}$ are concatenated with the intention prompt token $\mathbf{P}_{intent}$ and forms two sequence $\mathbf{Z}_{in} = \left [ \mathbf{P}_{intention}, \mathbf{V} \right ]$ and  $\mathbf{\hat{Z}} = \left [ \mathbf{P}_{intention}, \mathbf{\hat{V}} \right ]$. Then, the two sequences are fed into the same frozen LLM backbone, which is identical to the one deployed at the token receiver. From this shared model, we select a subset of intermediate layers as Semantic Evaluation (SE) layers, denoted as $\mathcal{L}=\left \{ l_1,l_2,\dots,l_K \right \} $. The outputs of these layers are considered as semantic representations. For each SE layer $l_k \in \mathcal{L}$, we apply mean pooling over the token dimension of its hidden states to obtain a global semantic vector
\begin{equation}
	\mathbf{h}^{(k)} = \frac{1}{T}\sum_{t=1}^{T}\mathcal{M}^{(l_k)}(\mathbf{Z}_{in})_t, \quad \mathbf{\hat{h}}^{(k)} = \frac{1}{T}\sum_{t=1}^{T}\mathcal{M}^{(l_k)}(\mathbf{\hat{Z}})_t,  
\end{equation} 
where $T=M+N$ is the total sequence length, and $\mathcal{M}^{(l_k)}(\cdot)_t$ denotes the hidden state of the $t$-th token at layer $l_k$.

We then compute the semantic cosine similarity (SeCS) for each layer as 
\begin{equation}
	\mathrm{SeCS}^{(k)} = \frac{\left \langle \mathbf{h}^{(k)},\mathbf{\hat{h}}^{(k)} \right \rangle }{\left \| \mathbf{h}^{(k)} \right \| \cdot \left \| \mathbf{\hat{h}}^{(k)} \right \|}.
\end{equation}
The overall semantic loss is defined as a weighted sum over all SE layers
\begin{equation}
	\mathcal{L}_{sem}=\sum_{k=1}^{K}i_k(1-\mathrm{SeCS}^{(k)}),
\end{equation}
where $i_k \ge 0$ are manually defined weights.

During the pre-training stage, the wireless channel is modeled as a differentiable surrogate, enabling efficient end-to-end optimization. The transmitted symbol sequence is corrupted by noise conditioned on the instantaneous SNR $\gamma$. To ensure stable gradient propagation, the noise is detached from the computational graph during backpropagation. Thus, gradients flow directly from the decoder input to the encoder output. This practice, widely adopted in deep joint source-channel coding \cite{9414037}, balances training stability with semantic robustness under stochastic distortions.

\subsubsection{End-to-End Fine-Tuning with LLM Backbone}
Building upon the semantically robust JTCM encoder and decoder modules obtained from the pre-training stage, we proceed to a task-specific fine-tuning stage that jointly optimizes the receiver-side LLM and the JTCM modules in an end-to-end manner. Specifically, the LoRA (Low Rank Adaptation) method is used to train the LLM during this phase by equipping adapter modules into the query and value projections of every multi-head attention layer, making the training cost-efficient. 

Overall, this training scheme serves three purposes. First, it enables efficient adaptation to the target task without modifying the original foundation model. Secondly, it refines the LLM's ability to interpret the concatenated input of the recovered visual tokens $\mathbf{\hat{V}}$ and the detailed task description prompt $\mathbf{P}_{task}$, achieving more precise task-level alignment. Thirdly, it steers the model's generation toward the required output format of the target downstream tasks.

During training, the receiver processes the channel-distorted symbols through the JTCM decoder to obtain $\mathbf{Z}_{rec} = \left [ \mathbf{P}_{task}, \mathbf{\hat{V}} \right ]$, which is fed into the LoRA-augmented LLM. The model then performs autoregressive next-token prediction, producing a sequence of output tokens $\left \{ \hat{t_1},\hat{t_2},\dots,\hat{t_T} \right \}$. We compute the standard teacher-forced cross-entropy loss against the ground-truth answer token sequence $\left \{ t_1,t_2,\dots,t_T \right \}$
\begin{equation}
	\mathcal{L}_{task}=-\sum_{\tau=1}^{T}\log P(t_\tau \mid t_{<\tau},\mathbf{Z}_{rec};\Theta_{\mathrm{LoRA}},\Theta_{\mathrm{JTCM}}),
\end{equation}
where $\Theta_{\mathrm{LoRA}}$ denotes the LoRA parameters and $\Theta_{\mathrm{JTCM}}$ includes both encoder and decoder weights. Crucially, gradients are backpropagated through the entire network, enabling the joint optimization of token compression, transmission, and task execution.

\section{Experiments} \label{experiments}
\subsection{Experiments Settings and Dataset Generation}
\begin{figure}
	\centering
	\subfigure[A sample image and questions from CLEVR.]{\label{phasechange}
		\includegraphics[width=0.3\textwidth]{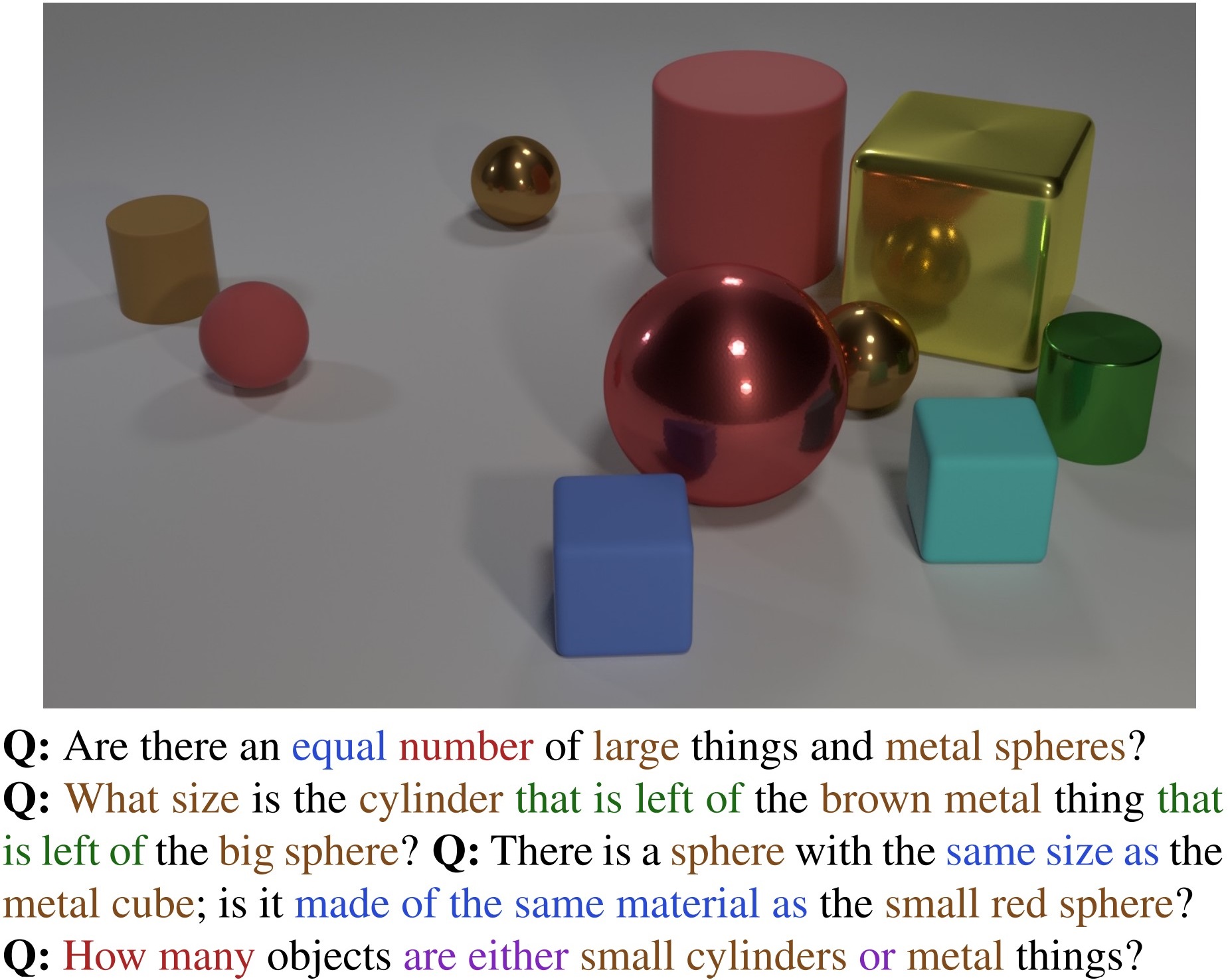}
	}
	\centering
	\subfigure[Examples from the GQA dataset for visual reasoning and compositional question answering.] {\label{phasesimulation_tanh}
		\includegraphics[width=0.3\textwidth]{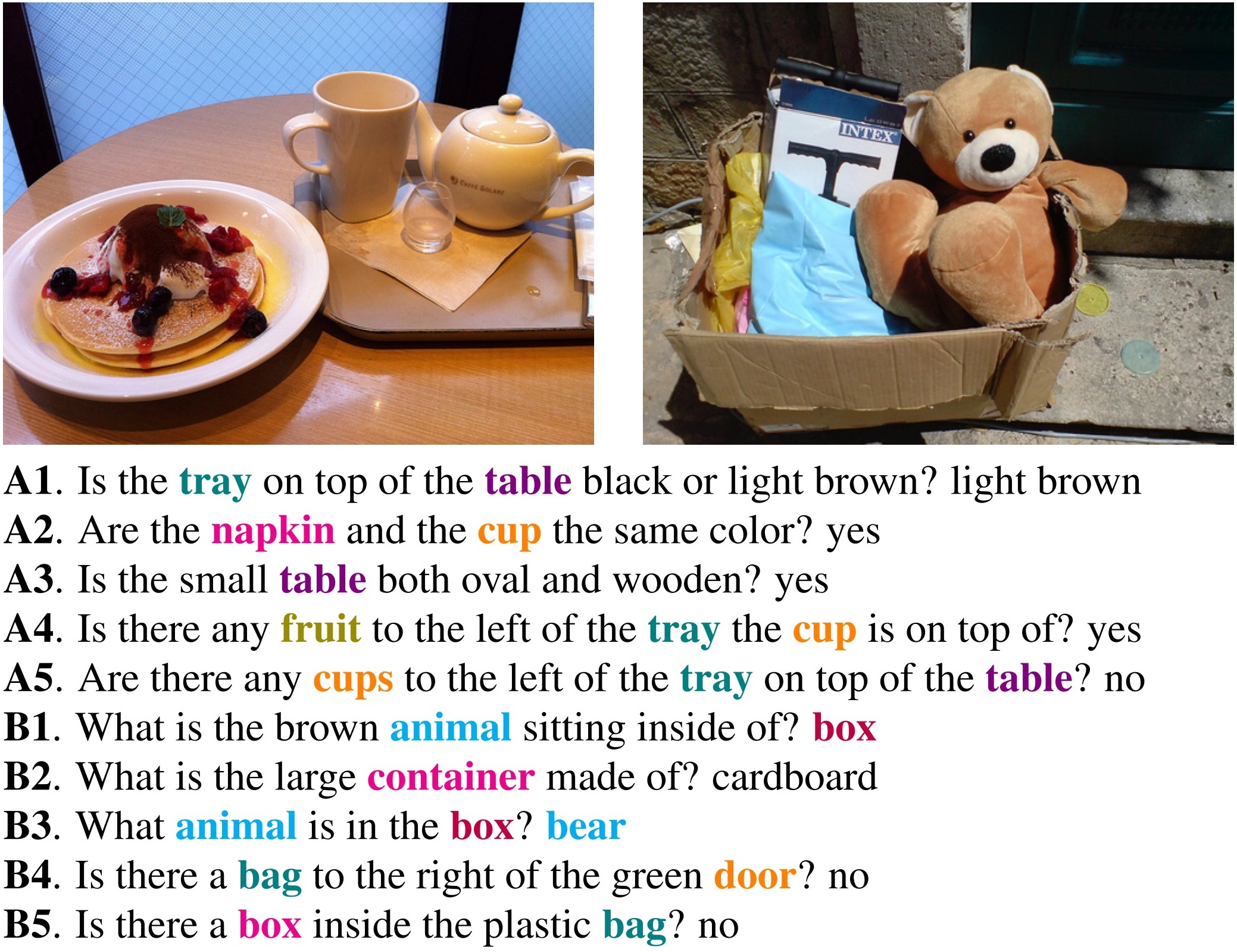}
	}
	\vspace{-.2cm}
	\caption{Illustration of the datasets used in the experiment part.}
	\vspace{-.5em}
	\label{datasets}
	\vspace{-.5cm}
\end{figure}

We adopt visual question answering as a representative multi-modal downstream task to evaluate the proposed method. Specifically, as shown in Fig. \ref{datasets}, we select CLEVR and GQA as our training and evaluation datasets. These datasets are chosen for two main reasons. One is that both CLEVR and GQA feature programmatically generated questions accompanied by rich scene graphs and metadata (e.g., object color, shape, spatial relationships). This enables controlled construction of task intent prompts, thereby aligning with the semantic abstraction capability targeted in our pre-training stage. The other is that both benchmarks emphasize compositional reasoning and multi-step inference, making them ideal for evaluating task-oriented semantic communication systems that aim to preserve not just perceptual fidelity but also reasoning-ready semantic structure during transmission.

\begin{figure}
	\vspace{-.5cm}
	\centering
	\includegraphics[width=0.3\textwidth]{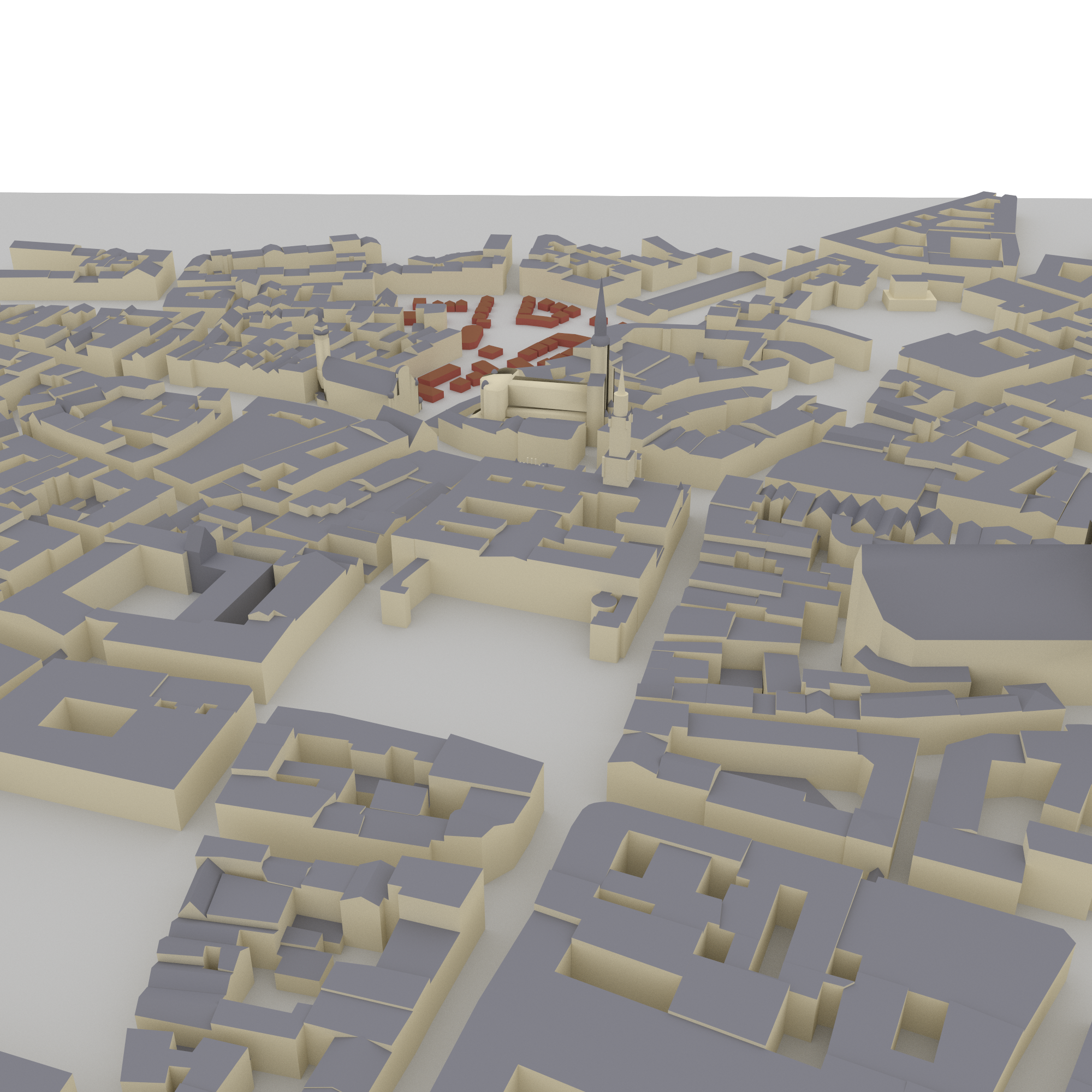}
	\caption{The 3D model of the scene used to collect channel samples.}
	\label{scene}
	\vspace{-.5cm}
\end{figure}

To evaluate the practical deployability of our system under realistic propagation conditions, we generate a ray-tracing channel dataset using the NVIDIA Sionna framework with the Munich urban map, as shown in Fig. \ref{scene}. We assume both transmitter and receiver are equipped with uniform linear arrays (ULAs) of $4$ antennas each. The system operates over $16$ orthogonal subcarriers, centered at $3.5$GHz with a subcarrier spacing of $100$MHz. A total $10000$ channel samples are collected for evaluation. Additive white Gaussian noise (AWGN) is added at the receiver side according to the target SNR. 

\subsection{Benchmarks}
To comprehensively evaluate the proposed methods, we compare them against a diverse set of baselines that span traditional communication, deep joint source-channel coding, and emerging semantic communication paradigms.
\begin{itemize}
	\item \textbf{Benchmark 1}: Images are first compresed using JPEG, then transmitted over 16QAM-modulated symbols. At the receiver, the image is reconstructed and fed into a frozen LLaVA-1.5-7B model for VQA.
	\item \textbf{Benchmark 2}: Reconstructing the original image using DeepJSCC with fixed latent feature dimension and using the reconstructed image as input to the LLaVA-1.5-7B model.
	\item \textbf{Benchmark 3}: The transmitter uses LLaVA-1.5-7B to generate a textual description of the image. The receiver directly uses this description with LLama2-7B-Instruct to answer the questions.
	\item \textbf{Benchmark 4}: The complete sequence of visual tokens (from ViT) and task prompt is passed to the receiver LLM without any channel distortion or bandwidth constraint. This serves as an upper bound on achievable task accuracy.
\end{itemize}

\subsection{Experimental Results}
\begin{figure}
	\centering
	\includegraphics[width=0.4\textwidth]{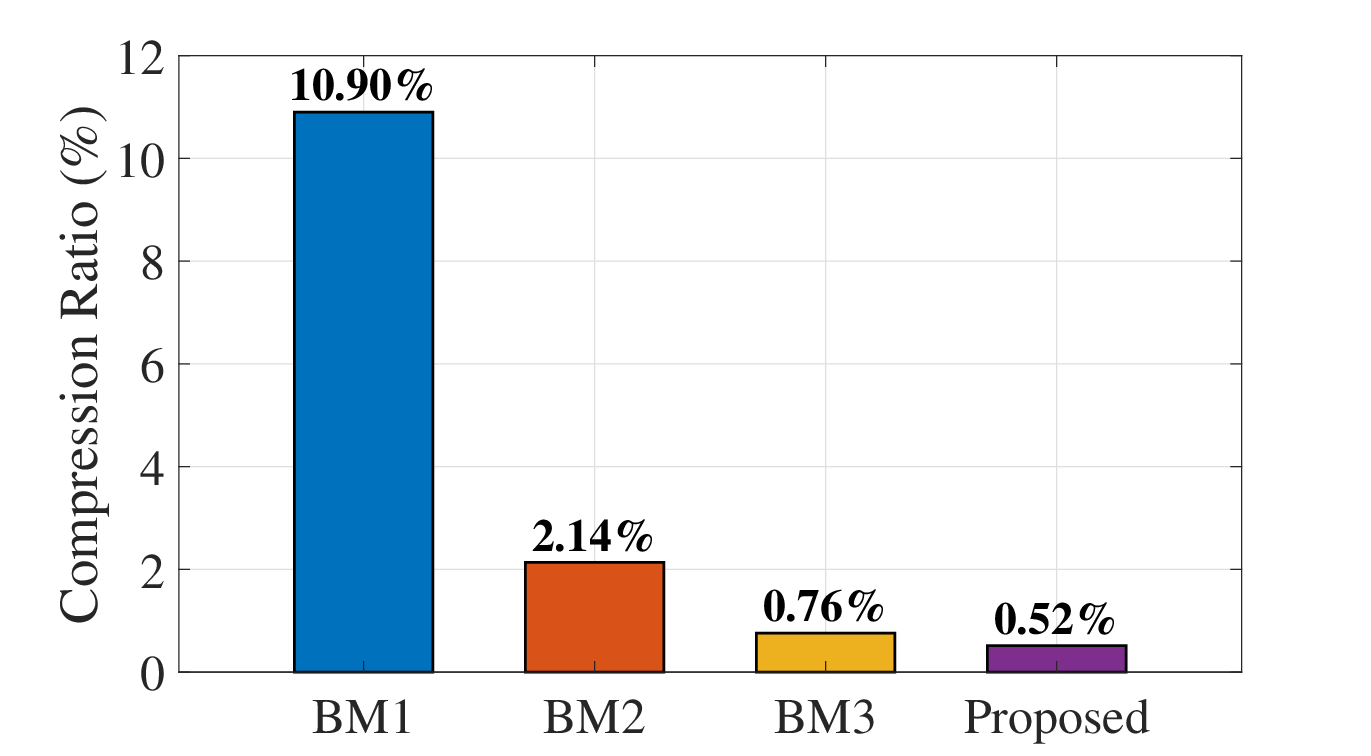}
	\vspace{-.4cm}
	\caption{Compression ratio relative to the original images.}
	\label{compression_ratio}
	\vspace{-.5cm}
\end{figure}

Fig. \ref{compression_ratio} illustrates the symbol compression ratios of various transmission schemes relative to the baseline. The baseline scheme transmits the raw RGB pixel bits of the original image using 16-QAM modulation. In Benchmark 2, the latent representation from a JSCC encoder is quantized with 8 bits per dimension and transmitted via 16-QAM modulation. Benchmark 3 encodes the textual description using UTF-8 and also employs 16-QAM modulation. Our proposed method requires only $2056$ channel symbols. Experimental results demonstrate that the proposed semantic communication approach significantly reduces over-the-air signaling overhead at the symbol level.
	
\begin{figure}
	\centering
	\subfigure[CLEVR dataset]{
		\begin{minipage}[c]{0.24\textwidth}
			\centering
			\includegraphics[width=1\textwidth]{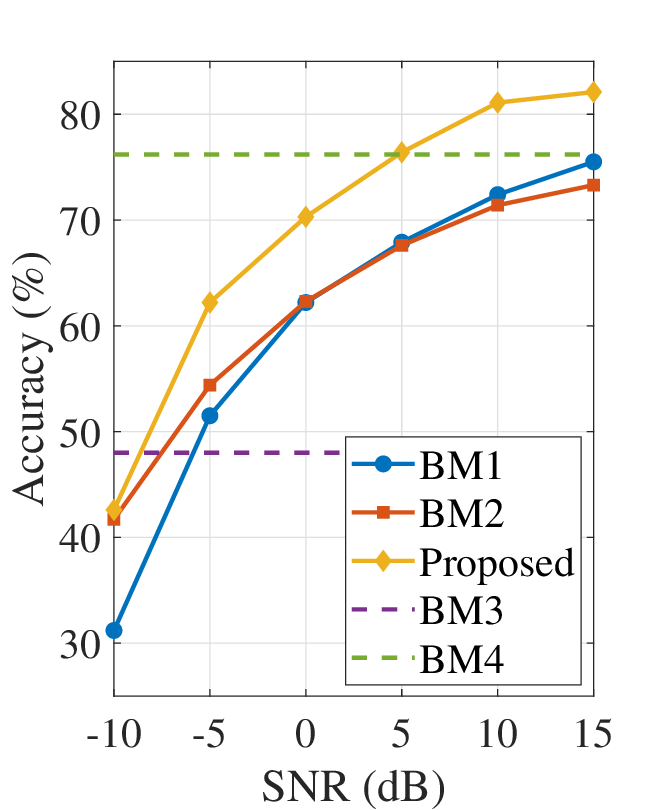}\vspace{-1em}
			\label{clevr_1}
		\end{minipage}
	}
	\hspace{-0.6cm}
	\subfigure[GQA dataset]{
		\begin{minipage}[c]{0.24\textwidth}
			\centering
			\includegraphics[width=1\textwidth]{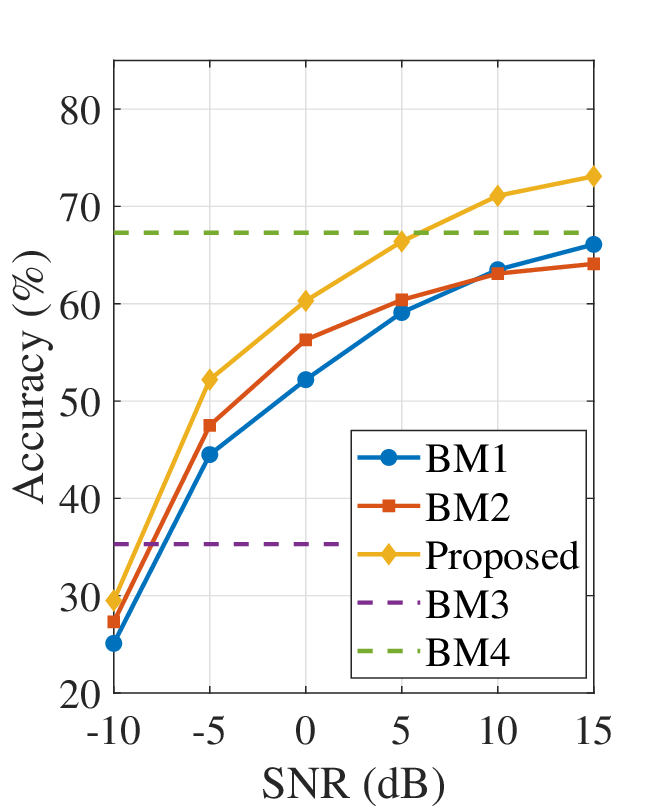}\vspace{-1em}
			\label{gqa_1}
		\end{minipage}
	}
	\vspace{-.4cm}
	\caption{The accuracy of downstream tasks across different methods with varying transmission SNR.}
	\label{exp1}
	\vspace{-.5cm}
\end{figure}

Fig. \ref{exp1} compares the downstream task accuracy of different methods across a range of SNR levels, revealing consistent trends on both datasets. Thanks to end-to-end training jointly optimized with the downstream tasks, the proposed method consistently outperforms all benchmarks at every SNR. In contrast, existing DeepJSCC approaches, which are primarily designed for raw image reconstruction, exhibit advantages only in low-SNR regimes. Benchmark 3, which relies on natural language descriptions, performs poorly on both datasets, likely due to the inherent difficulty of precisely conveying image-related reasoning tasks through text alone. Although Benchmark 1, which transmits compressed images, achieves reasonable performance under clean conditions, it suffers significant degradation in the presence of channel distortion. These results highlight the effectiveness and robustness of the proposed semantic communication framework.

\begin{figure}
	\centering
	\subfigure[CLEVR dataset]{
		\begin{minipage}[c]{0.24\textwidth}
			\centering
			\includegraphics[width=1\textwidth]{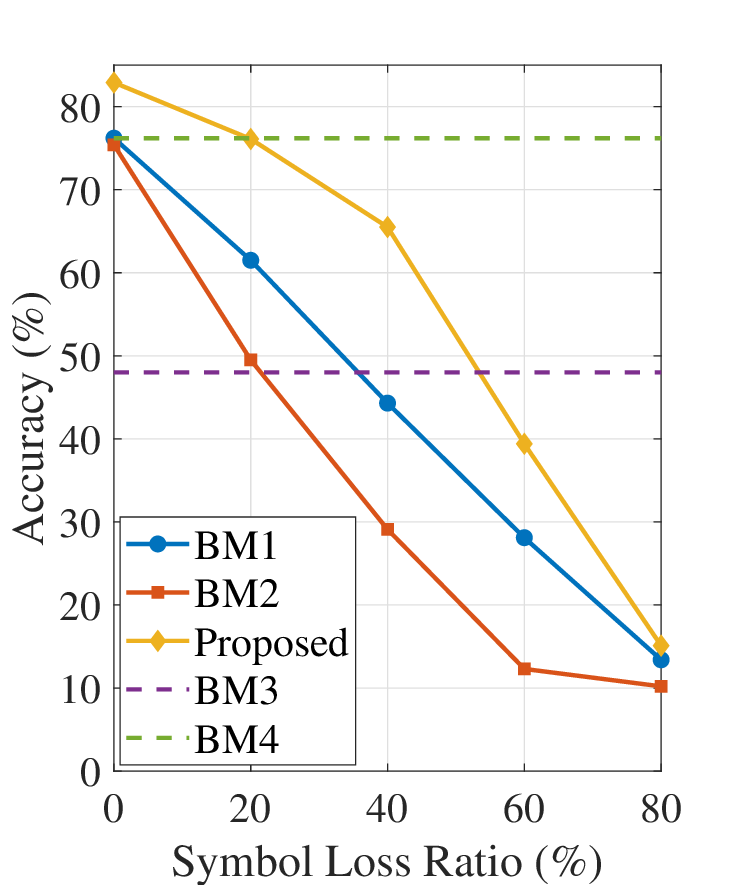}\vspace{-1em}
			\label{clevr_1}
		\end{minipage}
	}
	\hspace{-0.6cm}
	\subfigure[GQA dataset]{
		\begin{minipage}[c]{0.24\textwidth}
			\centering
			\includegraphics[width=1\textwidth]{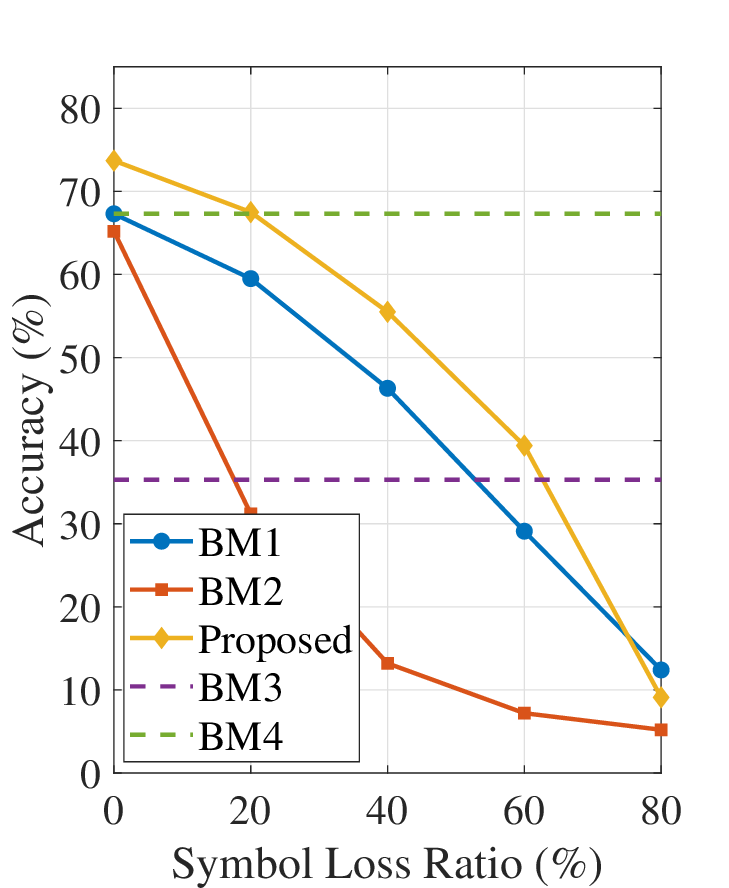}\vspace{-1em}
			\label{gqa_1}
		\end{minipage}
	}
	\vspace{-.4cm}
	\caption{The accuracy of downstream tasks across different methods with varying symbol loss ratio.}
	\label{exp2}
	\vspace{-.5cm}
\end{figure}

Fig. \ref{exp2} presents the performance of different methods under varying degrees of symbol loss. In the experiments, missing symbols in all benchmarks are replaced with the mean of the remaining symbols. Thanks to the structural properties of semantic tokens, the proposed method exhibits strong robustness to symbol loss, maintaining high task accuracy even when a non-negligible fraction of symbols is missing. In contrast, the scheme that transmits raw image pixels shows an almost linear degradation in performance as symbol loss increases, consistently underperforming our approach. Moreover, the DeepJSCC-based method suffers severe performance collapse under symbol loss, as its latent representations are highly sensitive to missing information. These results further confirm the robustness of the proposed semantic communication framework against channel impairments.

\section{Conclusions} \label{conclusion}
This paper proposed a pioneering paradigm for agent-centric token communication. By unifying semantic extraction, channel adaptation, and task execution into an end-to-end trainable pipeline, the proposed paradigm achieves efficient transmission using compact, task-relevant semantic token symbols. The uniquely designed token transmitter and receiver operate in synergy, with the transmitter distilling multimodal inputs into a minimal set of task-optimized semantic symbols, and the receiver utilizing a lightweight, task-aware large language model to generate structured responses directly from the distorted channel output. The experimental outcomes affirm the efficacy and robustness of the proposed methodology, paving the way for truly intelligent, task-oriented wireless agents.

	\bibliographystyle{IEEEtran}
	\bibliography{bibfile}
\end{document}